\documentclass[aps,pra,reprint,twocolumn,floatfix]{revtex4-2}
\usepackage{amsfonts, amsmath, amssymb}
\usepackage{graphicx}
\usepackage{dcolumn}
\usepackage{bm}
\usepackage[normalem]{ulem}
\usepackage{color}
\usepackage{dsfont}
\usepackage{siunitx}
\usepackage[utf8]{inputenc}
\usepackage[colorlinks,citecolor=blue,linkcolor=blue,urlcolor=blue]{hyperref}
\usepackage{footmisc}
\usepackage{adjustbox}
\usepackage{tikz}	
\usetikzlibrary{backgrounds,fit,decorations.pathreplacing}  
\usepackage{enumerate}

\newcommand{\ket}[1]{\ensuremath{\left|#1\right\rangle}} 
\usepackage{braket}
\usepackage{physics}
\usepackage{ulem}
\usepackage[caption=false]{subfig}
\usepackage{color}
\usepackage{xcolor}
\usepackage{soul}
\usepackage{mathtools}
\usepackage{float}
\newcommand{\be}{\begin{equation}}
\newcommand{\ee}{\end{equation}}
\newcommand{\ba}{\begin{eqnarray}}
\newcommand{\ea}{\end{eqnarray}}

\include{physics}

\newcommand{\pidx}[1]{{\mbox{\tiny $(#1)$}}}

{\left\lbrace\begin{array}{@{}l@{}}}%
	{\end{array}\right.}

\begin{document}
	
	\title{
		Continuous-time dynamics and error scaling of \\ noisy highly entangling quantum circuits}

	\begin{abstract}
		We investigate the continuous-time dynamics of highly-entangling intermediate-scale quantum circuits in the presence of dissipation and decoherence. By compressing the Hilbert space to a time-dependent ``corner" subspace that supports faithful representations of the density matrix, we simulate a noisy quantum Fourier transform processor with up to 21 qubits. Our method is efficient to compute with a controllable accuracy the time evolution of intermediate-scale open quantum systems with moderate entropy, while taking into account microscopic dissipative processes rather than relying on digital error models. The circuit size reached in our simulations allows to extract the scaling behavior of error propagation with the dissipation rates and the number of qubits. Moreover, we show that depending on the dissipative mechanisms at play, the choice of the input state has a strong impact on the performance of the quantum algorithm.
		
	\end{abstract}

	\date{\today}
	\author{Kaelan Donatella}
	\thanks{These two authors contributed equally.}
	\affiliation{Universit\'e de Paris, CNRS, Laboratoire Mat\'eriaux et Ph\'enom\`enes Quantiques, F-75013 Paris, France}

	\author{Zakari Denis}
	\thanks{These two authors contributed equally.}
	\affiliation{Universit\'e de Paris, CNRS, Laboratoire Mat\'eriaux et Ph\'enom\`enes Quantiques, F-75013 Paris, France}

	\author{Alexandre Le Boit\'e} \affiliation{Universit\'e de Paris, CNRS, Laboratoire Mat\'eriaux et Ph\'enom\`enes Quantiques, F-75013 Paris, France}

	\author{Cristiano Ciuti}
	\affiliation{Universit\'e de Paris, CNRS, Laboratoire Mat\'eriaux et Ph\'enom\`enes Quantiques, F-75013 Paris, France}
	
	\maketitle
	
	\section{Introduction}
	The tremendous advances on the control of artificial quantum systems, such as superconducting Josephson qubits \cite{Blais2020} and trapped ions \cite{trapped_ions}, are allowing dramatic progress toward the realization of devices for quantum computation \cite{Arute2019, Zhongeabe8770}.  We have reached the noisy intermediate-scale quantum (NISQ) era \cite{Preskill2018quantumcomputingin}, where error-correction is not yet possible due to daunting overheads \cite{devitt_quantum_2013}, but where quantum advantage might be already exploited for applications in quantum chemistry \cite{Lanyon2010}, optimization \cite{Biamonte2017} and even finance \cite{ORUS2019100028}. It is therefore of crucial importance to precisely understand the effects of both incoherent and coherent sources of noise on quantum algorithms~\cite{Martinis2015,Harper2020,deutsch_2020, Cross2019}. To meet these challenges, there is a strong need for accurate numerical simulations of quantum hardware on classical computers~\cite{Aharonov1996, Aaronson2016, Harrow2003,koch2020,LaRose2019overviewcomparison}. 
	In particular, the application of tensor network methods to quantum circuit simulation has been shown to be effective to model circuits with a limited degree of entanglement~\cite{Vidal2003,Zhou_2020,Noh_2020,Napp2020,Oh2021}. 
	
	Most existing simulators of quantum hardware consider local and digital error models \cite{devitt_quantum_2013,Jones2019, Arute2019,gottesman_2014}, that consist in extending the quantum circuit model to noisy algorithms by including noise gates applied after each unitary gate.
	Although in close proximity with classical error models, these two approximations do not necessarily hold, especially for highly-entangling circuits~\cite{weinstein_quantum_2004}, and remain a challenge in quantum error correction \cite{PhysRevLett.95.230503}. To take into account realistic sources of noise for highly-entangling circuits one should revert to a continuous-time description, where noise is taken into account continuously during the dynamics associated with the quantum gates. If one neglects non-Markovian effects, this can be performed in the framework of the Lindblad master equation \cite{breuer_theory_nodate}. However, such a description is numerically expensive; for a chain of $L$ qubits with Hilbert space dimension $N=2^L$, a density matrix of size $N^2$ must be evolved. Several proposals to reduce the complexity of the task that do not limit entanglement exist \cite{weimer2021,Weimer2015, Casteels2016,Kshetrimayum2017,Biella2018,kilda2021,Nagy2018,Nagy2019,Vicentini2019,Hartmann2019,Yoshioka2019}, such as the Monte-Carlo wavefunction method \cite{Carmichael1993,Molmer:93, Plenio1998} that reduces the problem to evolving many wavefunctions. However, their number is not known in general \cite{daley2014traj} and, in the case of weak dissipation, the method can quickly become equivalent to a full integration of the master equation as a greater amount of trajectories are needed to reach convergence. In recent years, there has been a growing interest in the idea that for a certain class of low-entropy systems, a limited number of states, belonging to a so-called ``corner" subspace, can efficiently and faithfully represent the density matrix  \cite{finazzi2015corner, rouchon,chen2021,fast_prr2020}. Since quantum processors are conceived to be weakly dissipative and with low entropy, they belong to this class. Stabilized arrays \cite{Ma2019, Lebreuilly2017}, cat qubit systems \cite{Guillaud2019} and quantum hardware with state-of-the-art dissipation rates \cite{Preskill2018quantumcomputingin,Arute2019} belong to this category.
	
	\begin{figure*}[ht!]
		\centering
		\includegraphics[width=\linewidth]{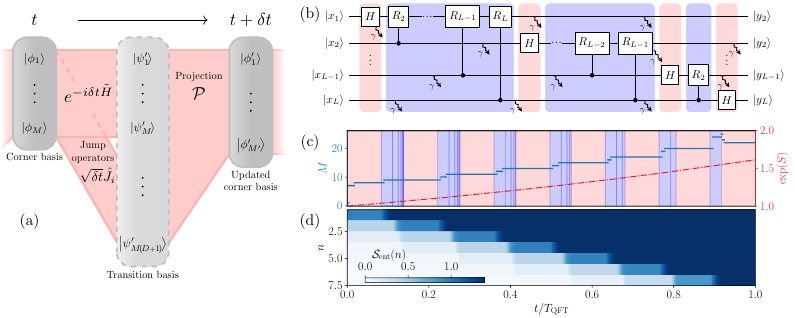}
		\caption{(a) Sketch depicting one iteration of the time-dependent corner-space method. 
			(b) Quantum circuit representing the QFT in the presence of dissipation.
			(c) Continuous-time evolution of the exponential of the von Neumann entropy $S$ (dash-dotted) with the $t=0$ input state $\ket{\psi_0} = \mathrm{QFT}^{-1} \ket{\mathrm{GHZ}}$ defined in the text, for $L=8$ qubits and $\gamma/\delta = 10^{-3}$. The corner-space dimension $M(t)$ is also plotted  (dash-dotted line). Temporal intervals corresponding to Hadamard and controlled phase gates (see text for more details) are indicated by lighter (red) and darker (blue) background colors, respectively. (d) Temporal build-up of the entanglement entropy function $\mathcal{S}_{\mathrm{ent}}(n) \equiv S(\tr_{1,\ldots, n}\hat{\rho})$, for all possible bipartitions of the qubit register having the form $\lbrace \lbrace 1,\ldots,n\rbrace, \lbrace n+1,\ldots, L\rbrace\rbrace$, as the different gates of the QFT are performed.
		}
		\label{fig:qft_evol}
	\end{figure*}
	
	In this paper, we investigate the continuous-time evolution of noisy intermediate-scale quantum circuits. We develop a time-dependent corner-space method with no restriction on the degree of entanglement, circuit connectivity, physical dimension or noise correlations and provide results with fully controllable accuracy. We focus our paper on the investigation of the role of dissipation and decoherence on the quantum Fourier transform (QFT).  This is an essential and highly-entangling quantum circuit at the heart of the Shor algorithm \cite{shor_algo}, quantum phase estimation \cite{phase_estimation} and many algorithms related to the hidden subgroup problem \cite{ETTINGER200443}. We demonstrate the capabilities of our method by simulating the dissipative QFT up to 21 qubits with a high and controlled accuracy, with a speedup of at least three orders of magnitude with respect to a full integration of the master equation. We show that the infidelity of the output state of the dissipative QFT with respect to the output state of a unitary QFT surprisingly scales polynomially with the system size, with an exponent that does not depend on the dissipation rate. 
	Furthermore, we explore the impact of different dissipative mechanisms on the fidelity and study how the initial state affects the performance of the quantum computation.
	
	\section{Time-dependent corner-space method\label{sec:2}} Let us consider an open quantum system whose dynamics is governed by the following Lindblad master equation \cite{breuer_theory_nodate}:
	\begin{equation}\label{eq:1}
	\partial_t\hat{\rho} = -\text{i}[\hat{H},\hat{\rho}] + \sum_{i=1}^D\big(\hat{J}_i\hat{\rho}\hat{J}_i^\dagger - \frac{1}{2}\lbrace\hat{J}_i^\dagger\hat{J}_i,\hat{\rho}\rbrace\big),
	\end{equation}
	where $\hat{H}$ is the system Hamiltonian ($\hbar = 1$) acting on a Hilbert space $\mathcal{H}$ of dimension $N$, and $\hat{J}_i$ is the $i$-th jump operator. At any time $t$, the solution $\hat{\rho}$ may be approximated by:
	\begin{equation}\label{eq:2}
	\hat{\rho}(t) \simeq \sum_{k=1}^{M(t)} p_k(t)\ketbra{\phi_k(t)}{\phi_k(t)}, 
	\end{equation}
	where $p_k(t)$ are the $M(t)$ largest eigenvalues of $\hat{\rho}$ at the time $t$ [we order the eigenvalues in such a way that $p_k(t) \geq p_{k+1}(t)$, $\forall k$] and the $\ket{\phi_k(t)}$'s their associated eigenvectors. By construction, the controlled truncation error of such an approximation is strictly decreasing in $M$ and quantified by $\epsilon_M = 1 - \sum_{k=1}^M p_k$ so that the decomposition becomes exact for $M(t) = r(t)$ with $r(t)$ denoting the rank of $\hat{\rho}(t)$, equivalent to the $\alpha = 0$ Rényi entropy \cite{beck_schogl_1993}. Therefore, in a wide class of low-entropy systems including most platforms relevant for quantum computing, $\hat{\rho}$ is very well approximated by $M\ll N$ basis vectors, and even by $M\gtrsim 1$ for close to pure states. Henceforth, $M$ will be referred to as the \emph{corner dimension}. The accuracy of the calculations will be controlled by a fixed maximum error $\epsilon$ with $\epsilon_M\leq\epsilon$ enforced at any time.
	
	All the information of the density matrix is carried by a set of weighted corner basis vectors of the form $\sqrt{p_k}\ket{\phi_k}$. In some arbitrary computational basis $\lbrace\ket{n}\rbrace_{n=1}^{N}$, these can be represented by a $N\times M$ matrix with elements $C_{nk}(t) = \sqrt{p_k(t)}\braket{n}{\phi_k(t)}$. From Eq.~\eqref{eq:2} one has:
	
	\begin{equation}\label{eq:3}
	\hat{\rho}(t) = \hat{C}(t)\hat{C}^\dagger(t).
	\end{equation}
	
	\begin{figure*}[t!]
		\centering
		\includegraphics[width=\textwidth]{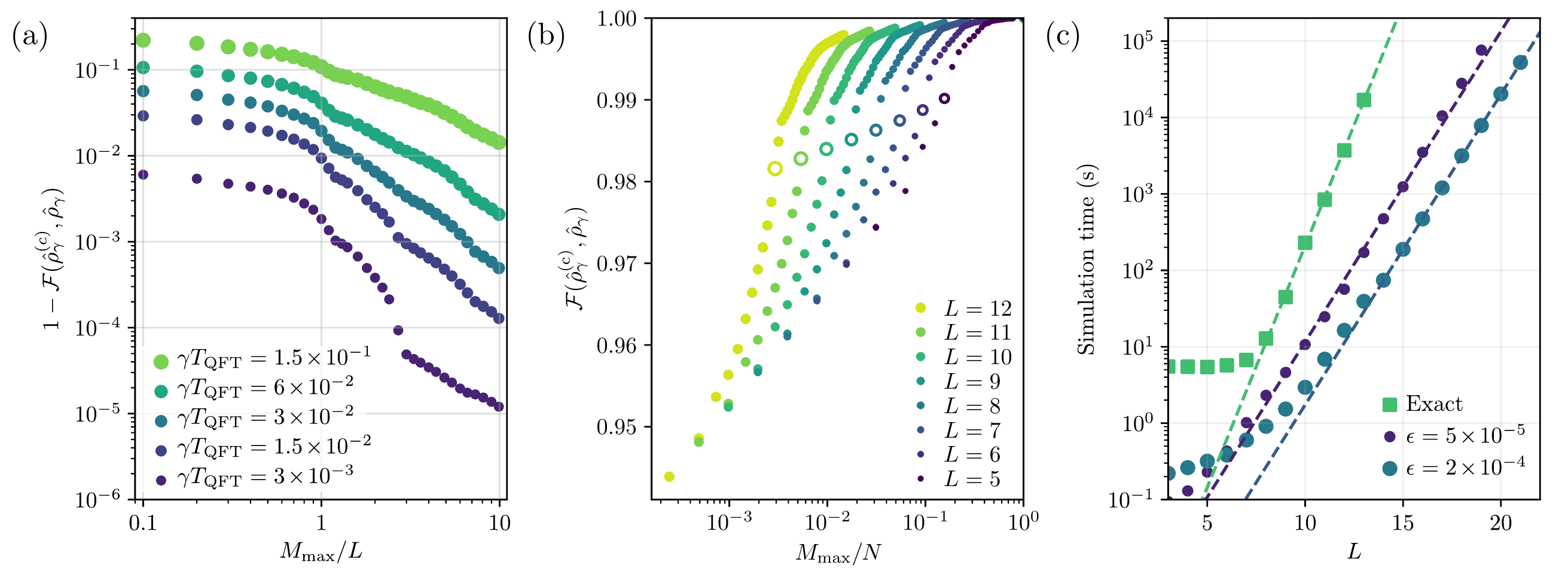}
		\caption{ (a) Infidelity between the output density matrix $\hat{\rho}_\gamma^\pidx{c}$ calculated with the corner-space method and the output $\hat{\rho}_\gamma$ obtained via an exact integration as a function of the maximum corner dimension $M_\mathrm{max}$ for different values of $\gamma T_{\text{QFT}}$ and a fixed number of qubits $L = 10$. (b) Fidelity as a function of $M_{\mathrm{max}}$ and $L$ for $\gamma T_\mathrm{QFT}=\num{2.5e-2}$. The parameter $N$ denotes the dimension of the total Hilbert space. Values corresponding to $M_\mathrm{max}=L$ are highlighted by hollow markers (we find a fidelity $\mathcal{F} \gtrsim 0.997$ for $M_\mathrm{max} \sim L\ln L$). The initial state is $\ket{\psi_{\mathrm{0}}}$ as in Fig.~\ref{fig:qft_evol} for all three panels. (c) Simulation time of the QFT algorithm in the presence of dissipation versus the number $L$ of qubits for the exact solution of the master equation (squared markers) and for the time-dependent corner-space method (circles) for two different values of the control parameter $\epsilon$. The dissipation rate is set by $\gamma T_{\text{QFT}} = \num{2.5e-2}$.
		}
		\label{fig:benchmarks}
	\end{figure*}
	
	The essential goal of this method is to efficiently perform the time-evolution of the low-dimensional weighted corner basis $\hat{C}$ without ever reconstructing $\hat{\rho}$. The evolution $\hat{C}(t)\mapsto\hat{C}(t+\delta t)$ over a small time step $\delta t$, schematically illustrated in Fig.~\ref{fig:qft_evol}(a), involves two computational operations: (i) calculation of the transition basis and (ii) dimensional reduction by projection onto the new principal components.
	
	Step (i): The weighted corner basis $\hat{C}(t)$ evolves into the weighted transition basis $\hat{T}(t+\delta t)$ as
	\begin{equation}\label{eq:4}
	\hat{\rho}(t+\delta t) = \sum_{i=0}^{D}\hat{K}_i\hat{\rho}(t)\hat{K}^\dagger_i = \hat{T}(t+\delta t)\hat{T}^\dagger(t+\delta t),
	\end{equation}
	where we used the Kraus map \cite{wiseman_milburn} equivalent to Eq.~\eqref{eq:1}, with $\hat{K}_0 = \mathrm{exp}(-\text{i}\delta t\tilde{H})$ and $\tilde{H} = \hat{H}-\frac{i}{2}{\textstyle\sum_{i=1}^D}\hat{J}_i^\dagger\hat{J}_i$ a non-Hermitian operator depending on the Hamiltonian and on the quantum jump operators. The other Kraus operators are $\hat{K}_i = \sqrt{\delta t}\hat{J}_i$. Note that by construction $\hat{T}(t+\delta t)$~\footnote{whose $m$th column is given by $\ket{\psi_m(t+\delta t)} = \sqrt{p_\mu}\hat{K}_\nu\ket{\phi_\mu(t)}$, with $\nu = (m-1) \divisionsymbol M(t)$ and $\mu = (m-1)\mod M(t) + 1$.} is a $N[M(t)(D+1)]$ matrix, where $D$ is the number of dissipation channels. Even though the Kraus operators are $N\times N$ matrices that have to be dealt with, they are, in general, extremely sparse matrices (see Appendix~\ref{sec:appA.2}).
	
	Step (ii): The transition matrix is now projected to a new weighted corner basis $\hat{C}(t+\delta t)$ of (lower) dimension $M(t+\delta t)$ via a new truncated eigendecomposition $\mathcal{P}$ of the form of Eq.~\eqref{eq:3}. Importantly, this is possible without ever reconstructing the full density matrix. Indeed, $\hat{\rho}(t+\delta t) = \hat{T}(t+\delta t)\hat{T}^\dagger(t+\delta t)$ (whose number of elements is equal to $N^2$) and $\hat{\sigma}(t+\delta t) = \hat{T}^\dagger(t+\delta t) \hat{T}(t+\delta t)$ (whose number of elements is $[M(t)(D+1)]^2$) share the same non-zero eigenvalues $p_k$ with eigenvectors $\ket{\phi_{k,\rho}(t+\delta t)}$ and $\ket{\phi_{k,\sigma}(t+\delta t)}$. These are related by the relation $\sqrt{p_k}\ket{\phi_{k,\rho}(t+\delta t)} = \hat{T}(t+\delta t)\ket{\phi_{k,\sigma}(t+\delta t)}$ \cite{gentle_svd}. The components of the decomposition can be judiciously truncated to retain the leading $M(t+\delta t)$ eigenvalues $p_k$, yielding an updated weighted corner basis $\hat{C}(t+\delta t)$, with the same structure as the initial one $\hat{C}(t)$. 
	
	In our implementation (see Appendix~\ref{sec:appA.5}), the coherent part of the evolution is integrated with high-order and stiffness-stable techniques, and the corner dimension $M(t)$ is dynamically adapted to match the desired accuracy, given by $\epsilon$. As a result, the presented method can deal with arbitrary Markovian noise models with no limit on the degree of entanglement and amounts to evolving $O(ML\, 2^L)$ closed systems (see Appendix~\ref{sec:appA.1}). Our method is thus particularly suited for high-entangling quantum circuits with moderate entropy as the dimension $M$ ultimately depends on the entropy.
	
	\section{Application to the noisy QFT} As a first application, we numerically simulate a noisy QFT on the circuit shown in Fig.~\ref{fig:qft_evol}(b) with up to $L = 21$ qubits. Quantum gates are executed via a continuous-time evolution defined by an appropriate master equation taking the form of Eq.~\eqref{eq:1}. 
	We work in a frame rotating at the frequency of the qubit transition. We model the noise as dissipation due to an environment at zero temperature described by jump operators of the form $\hat{J}_i = \sqrt{\gamma}\hat{\sigma}_i^-$, corresponding to local decay processes from the excited qubit state $\ket{\uparrow}_j \equiv \ket{0}$ to the lower energy qubit state  $\ket{\downarrow}_j \equiv \ket{1}$.   The Hadamard gate on qubit $i$ is achieved via the application of the Hamiltonian $\hat{H}^1_i = \frac{\delta}{2} \hat{\sigma}_i^y$ for a time duration equal to $\pi / 2 \delta$ ($\pi/2$-rotation along the $y$-axis of the qubit Bloch sphere) followed by the application of $\hat{H}^2_i = \frac{\delta}{2} \hat{\sigma}_i^z$ for a time $\pi / \delta$ ($\pi$-rotation along the $z$-axis of the Bloch sphere). A controlled phase gate of angle $\theta$ with control qubit $j$ and target qubit $k$ is performed by the Hamiltonian $\hat{H}_{j,k} = \frac{\delta}{2} \hat{\sigma}_j^z + \frac{\delta}{2} \hat{\sigma}_k^z - \frac{\delta}{2} \left(\hat{\sigma}_j^z\hat{\sigma}_k^z  + \hat{\mathds{1}} \right)$ for a time $\theta/\delta$. For simplicity, we assume sudden switching between gate Hamiltonians and do not include coherent errors, although both effects can be accurately described by our method. 
	
	We present an example of dynamics in Figs.~\ref{fig:qft_evol}(c) and \ref{fig:qft_evol}(d). As an initial input state, we choose $\ket{\psi_0} = \text{QFT}^{-1} \ket{\text{GHZ}}$~\footnote{This state can be explicitly written $\ket{\psi_0} = (1/\sqrt{2N})\sum_{n=0}^{N-1}(1 + e^{2i\pi n/N})\ket{n}$, with $\ket{n}$ the $n$th state of the computational basis.} with $\ket{\text{GHZ}} = \frac{1}{\sqrt{2}} \left(\ket{00...0} + \ket{11...1}\right)$~\cite{GHZ} in order to get a highly-entangled output state close to $\ket{\text{GHZ}}$ and demonstrate that entanglement is not a limiting factor for our method. In panel (c), the time evolution of the corner-space dimension $M$ as well as the exponential $\exp(S)$ of the von Neumann entropy $S(\hat{\rho})= \Tr{\hat{\rho}\ln{\hat{\rho}}}$ are shown.  Panel (d) shows the entanglement entropy $\mathcal{S}_{\mathrm{ent}}(n) \equiv S(\tr_{1,\ldots,n}\hat{\rho})$ calculated for bipartitions of the form $ \lbrace\lbrace 1,\ldots,n\rbrace, \lbrace n+1,\ldots,L\rbrace \rbrace$. The entanglement entropy is a measure of entanglement only for pure states~\cite{horodecki_2009}, however given the moderate entropy here, it gives a qualitative idea of the entanglement temporal build-up.
	
	\begin{figure}[t]
		\centering
		\includegraphics[width = \linewidth]{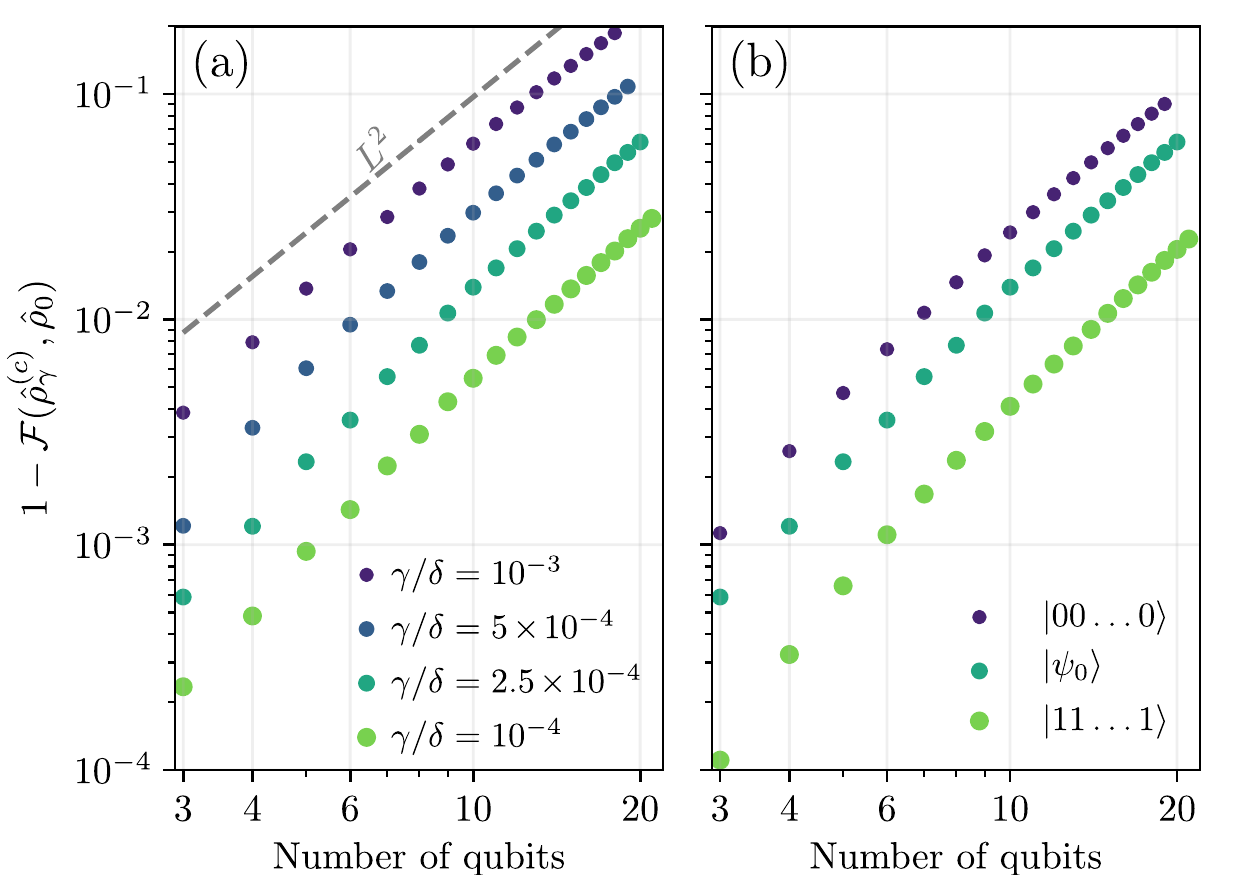}
		\caption{(a) Infidelity between the noisy output density matrix $\rho^{(c)}_{\gamma}$ and the noiseless output $ \hat{\rho}_0 = \ketbra{\mathrm{GHZ}}{\mathrm{GHZ}}$ obtained from the initial state $\ket{\psi_0}$ for different values of $\gamma / \delta$. The dashed line is a guide to the eye showing a growth $\propto L^2$. (b)  Infidelity for three different initial states, described in the legend, for the same dissipation rate $\gamma / \delta  = \num{2.5e-4}$.}
		\label{fig:fidelityvsL}
	\end{figure}
	
	The accuracy of our calculations has been benchmarked to the results of an exact integration of the master equation for small values of $L$, the numbers of qubits. 
	In what follows, $\hat{\rho}_\gamma$ and $\hat{\rho}_\gamma^\pidx{c}$ denote the output density matrices of the noisy QFT obtained via the exact integration and via the corner method, respectively. Instead, $\hat{\rho}_0$ denotes the outcome of the noiseless QFT, which is a pure state.
	Results of our benchmark are presented in Fig.~\ref{fig:benchmarks} for fixed values of $\gamma T_\mathrm{QFT}$, where $T_\mathrm{QFT}$ denotes the physical duration of the QFT operation. This ensures that the output infidelity with respect to $\hat{\rho}_0$ remains constant as the circuit size is increased. In particular, 
	Fig.~\ref{fig:benchmarks}(a) shows the infidelity $1 - \mathcal{F}(\hat{\rho}_\gamma^\pidx{c},\hat{\rho}_\gamma)$ as a function of the rescaled maximum corner dimension $M_{\mathrm{max}}/L$ for $L=10$. We see that for $M_{\mathrm{max}} \sim 10 L$ the exact results are excellently approximated by the time-dependent corner-space method for all the considered values of $\gamma T_{\mathrm{QFT}}$. The method still performs reasonably well for noise rates as high as $\gamma T_{\mathrm{QFT}} = \num{1.5e-1}$, where the fidelity to the output of the noiseless circuit is as low as $\mathcal{F}(\hat{\rho}_0,\hat{\rho}_\gamma) = 0.758$. In Fig.~\ref{fig:benchmarks}(b) we show the fidelity $\mathcal{F}(\hat{\rho}_\gamma^\pidx{c},\hat{\rho}_\gamma)$ for different values of $L$ as a function of $M_{\mathrm{max}} / N$. These results show that the advantage of our method over exact integration of the master equation increases with $L$. In particular, for a given value of the fidelity, $M$ grows with the system size as $L\ln L$ (see Appendix~\ref{sec:appA.4}). For $L = 12$, an excellent agreement of the corner-space method with the exact integration is already obtained for $M_{\mathrm{max}} / N = 10^{-2}$. Finally, in Fig.~\ref{fig:benchmarks}(c), we compare the computation time of the corner-space method  to the exact integration, for two different values of $\epsilon$~\footnote{Both the exact integration and corner-space calculations were carried out on a single six-core Intel Xeon E5-2609 v3 processor at $1.9$\,GHz}.  Our method presents an exponential speed-up with respect to the master equation integration which leads to a simulation that is faster by more than three orders of magnitude for $L \sim 15$. Moreover, tuning  $\epsilon$ from $\num{2e-4}$ down to $\num{5e-5}$  preserves the scaling of the simulation time with $L$.
	\begin{figure}[t!]
		\centering
		\includegraphics[width = \linewidth]{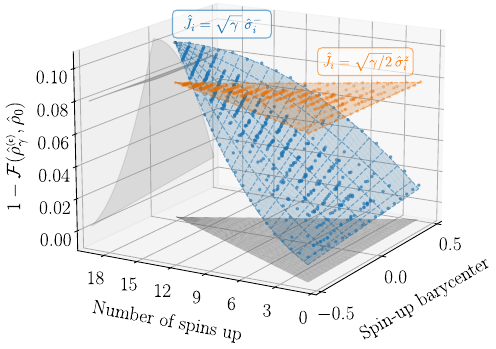}
		\caption{Infidelity for $L=18$ qubits with dissipation (dark blue) or pure dephasing (orange) for $517$ initial states randomly sampled from the canonical basis. The infidelity is plotted versus the total number $n_S$ of spins up in a given state and the spin-up barycenter (see the text). The infidelity in both cases can be fitted by $1 - \mathcal{F}(n_S,B) = a(n_S-n_{S_0})(B-B_0)+\mathcal{I}_0$. The surface for pure dephasing shows a negligible dependence on the initial state. Parameters: $\gamma/\delta = \num{3.7e-4}$ ($\gamma T_\mathrm{QFT} = \num{2.5e-2}$)  and $\epsilon = 10^{-4}$.}
		\label{fig:initstatedependence}
	\end{figure}    
	
	\section{Scaling laws and impact of initial states} We can now evaluate the impact of incoherent processes on intermediate-scale devices via a continuous-time description and determine the scaling of errors. In Fig.~\ref{fig:fidelityvsL}, we show the fidelity $\mathcal{F}(\hat{\rho}_\gamma^\pidx{c},\hat{\rho}_0)$ for up to $L = 21$ qubits (dimension $N = 2^{21} = \num[group-separator={,}]{2097152}$), for different values of $\gamma / \delta$. Here, we consider dissipation channels described by the jump operators  $\hat{J}_i = \sqrt{\gamma}\hat{\sigma}_i^- $. We find that the infidelity scales quadratically with the number of qubits $L$. Although this result may be understood qualitatively by considering that the total time of the algorithm is linear in $L$ and assuming a constant error rate, the fact that the scaling holds for different initial states is not trivial. For instance, starting from the $\ket{11...1}$ (all spins down) state, dissipation does not affect the $L-1$ last qubits during the first Hadamard gate and the following controlled phase gates, making the constant error assumption break down. Furthermore, as shown in Appendix~\ref{sec:appB}, errors depend on the applied gate, which are not all of the same duration.
	
	Another key property is the fidelity dependence on the initial state, crucial to redesign algorithms that rely preferentially on a certain class of states. In Fig.~\ref{fig:initstatedependence}, we address this question for the QFT by sampling initial states. Either energy relaxation produced by the jump operators $\hat{J}_i = \sqrt{\gamma}\hat{\sigma}_{i}^{-}$ or pure dephasing described by $\hat{J}_i = \sqrt{\gamma}\hat{\sigma}_i^z$ are considered. We have found that two simple parameters characterizing the initial state are crucial: the  total number $n_S$ of spins up and the spin-up ``barycenter" $B(\hat{\rho}) \equiv  (1/n_S)\sum_\ell \ell\Tr[\ketbra{\uparrow}{\uparrow}_\ell\hat{\rho}]$. Our findings show that, in the presence of energy relaxation, the fidelity of the noisy QFT decreases linearly with the number of spins up in the initial state. This is in stark contrast to the case of pure dephasing, which shows no significant dependence on $n_S$. The fidelity also exhibits a strong dependence on the spin barycenter. Indeed, energy relaxation affects excited states only and the circuit's Hadamard gates are applied one qubit at a time starting from the beginning of the chain. As a result, excited qubit states (spin up) close to the end of the chain are rotated down to the qubit Bloch sphere equator by the Hadamard gates later than those on the opposite end. Thus, they are globally more affected by dissipation.
	
	\section{Conclusion} 
	We investigated the role of dissipation and decoherence in noisy intermediate-scale quantum circuits. Focusing on the key algorithm of the QFT, we revealed the scaling behavior for the fidelity and explored its dependence on the initial state. To achieve this goal, we have introduced and demonstrated a numerical time-dependent corner-space method that performs a judicious compression of the Hilbert space to faithfully represent the system density matrix. The method is not limited by entanglement and is suitable for systems with moderate entropy. Furthermore, our approach could be combined with efficient representations of the corner-space wavefunctions, such as neural-network ansätze~\cite{carleo2017,carleo2018,schmitt2020,sharir2020}. These qualities make our approach ideally tailored for the NISQ era, providing a tool to improve our understanding of quantum hardware. The presented method can indeed be applied in many contexts related to quantum information: algorithm design for quantum feedback \cite{marquardt_2018}, machine learning for quantum control \cite{ZENG2020126886} and quantum error mitigation~\cite{sun2021}.
	\begin{acknowledgments}
		We acknowledge support from H2020-FETFLAG Project PhoQus (Project No. 820392) and ERC CoG NOMLI (Project No. 770933). 
	\end{acknowledgments} 

\appendix

\section{Computational details\label{sec:appA}}

\subsection{Complexity\label{sec:appA.1}}

The method presented above amounts to evolving $M(t)$ closed systems with $M(t)$ as the corner dimension. For a quantum computation, the initial state is pure, $M(t=0) = 1$. As shown in Fig.~\ref{fig:qft_evol}(c), the dimension $M(t)$ grows moderately in time. At every time step, the most demanding operation is related to the construction of the matrix $\hat{\sigma} = \hat{T}^\dagger\hat{T}$. This involves a number of operations of order $O(M^2 [D+1]^2 2^L)$. Indeed, the transition matrix $\hat{T}$ has a size $M(D+1) \times N$, where $N = 2^L$ denotes the dimension of the Hilbert space and $L$ denotes the size of the system under consideration. Note that the number of jump operators $D$ typically scales as $L$ for most relevant quantum computing platforms consisting of $L$ coupled units. The complexity of the method is thus of order $O(M^2 L^2 2^L)$. This represents an exponential reduction of the complexity with respect to that of a brute-force master equation integration, which is of order $O(4^L)$.

\subsection{Memory use\label{sec:appA.2}}

\begin{figure*}[!t]
    \centering
    \includegraphics[width=\textwidth]{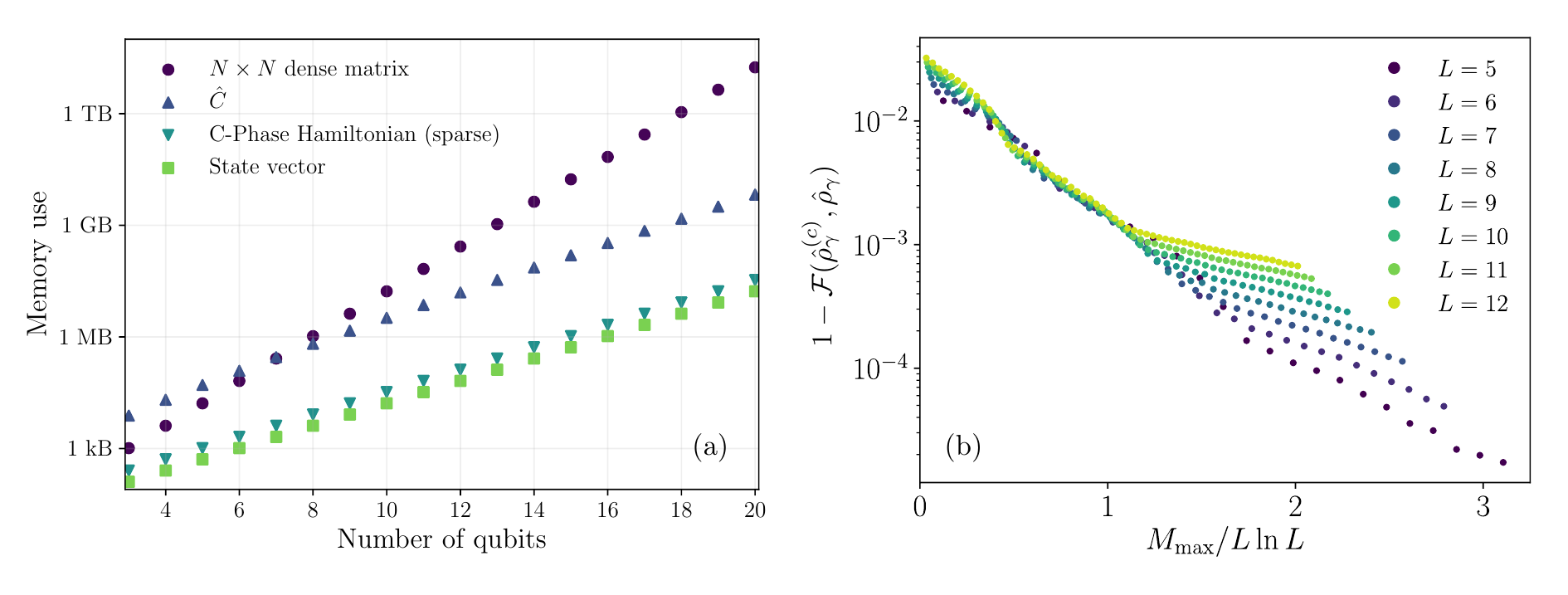}
    \caption{%
        (a) Memory use of different quantities. One can see that the gain in memory requirement from our method scales exponentially. Note that the Hamiltonian, being very sparse, is not the limitation. The largest stored object is the corner $\hat{C}$, that is composed of $M$ states. Here, as an example, we considered $M = 100$. (b) Infidelity between the dissipative output density matrix $\hat{\rho}^{\pidx{c}}_{\gamma}$ (calculated via the corner-space method)  and the exact integration density matrix $\hat{\rho}_{\gamma}$ as a  function of the maximum corner dimension $M_{\mathrm{max}}$ for an $L$-qubit circuit and $\gamma T_\mathrm{QFT}=\num{1.42e-2}$. The initial state is $\ket{\psi_0} = \mathrm{QFT}^{-1} \ket{\mathrm{GHZ}}$.
    }
    \label{fig:S1}
\end{figure*}

As pointed out in Sec.~\ref{sec:2}, the Kraus operators $\hat{K}_i$ are $N\times N$ matrices, the same size as the density matrix. However, the memory required to store them is by no means comparable because of their extreme sparsity. Indeed, jump operators corresponding to dissipative processes are typically single-body operators and thus as memory-consuming as state vectors. The largest Kraus operator is $\hat{K}_0$, which roughly corresponds to $\hat{H}$. The RAM needed to store it is shown in Fig.~\ref{fig:S1} together with that associated with other numerically relevant objects as a function of the number of qubits.
Another feature of our implementation is that the dimension of the corner basis $M(t)$ is dynamically adapted to match the maximum allowed error that we impose, thus optimizing the computing and memory resources. This is performed by means of memory-contiguous dynamically resizable arrays. This saves considerable computation time, as $M(t)$ grows in time when starting from a pure state [$M(0) = 1$].
\subsection{Efficient evaluation of relevant metrics}

The evaluation of relevant metrics in quantum information is challenging with quantum-trajectory approaches, such as the Monte Carlo wave function method~\cite{Molmer:93,Carmichael1993,Plenio1998} (MCWF). Such an approach relies on the evolution of $n$ stochastic trajectories $\lbrace\ket{\psi_i(t)}\rbrace_{i=1}^n$. The density matrix can then be reconstructed as $\hat{\rho}(t) = (1/n)\sum_{i=1}^n \ketbra{\psi_i(t)}$. To simulate a quantum computation, this has two downsides: first, for weakly dissipative systems ($\gamma T \ll 1$), most trajectories will experience no quantum jump and, thus, be identical. This means that most of the computing resources are wasted in performing a redundant task. Second, most relevant metrics involved in quantum-information processing, such as fidelity and entanglement measures, namely concurrence, negativity, or entanglement entropy~\cite{horodecki_2009}, require constructing explicitly the (dense) density matrix of the system and diagonalizing it. In practice, the latter, of complexity $O(N^3)$, is not feasible for systems larger than $\sim 15$ qubits. In contrast, the presented method yields explicitly both the eigenvalues $\lbrace p_k(t)\rbrace_{k=1}^M$ and the eigenvectors $\lbrace \ket{\phi_k(t)}\rbrace_{k=1}^M$ at every time step, with no need for additional calculations.

To give a concrete example, let us consider the evaluation of the fidelity between two arbitrary mixed states $\hat{\rho}$ and $\hat{\rho}'$ with ranks $M$ and $M'$, respectively, as given by
\begin{align*}
    \mathcal{F}(\hat{\rho},\hat{\rho}') &= \Tr[\sqrt{\sqrt{\hat{\rho}}\hat{\rho}'\sqrt{\hat{\rho}}}] \\
    &= \sum_{m=1}^M\bigl\langle\phi_m\bigr\rvert\Bigl\lbrace{\textstyle\sum_{k,m=1}^M}\ket{\phi_k}\mathcal{M}_{km}\bra{\phi_m}\Bigr\rbrace^{1/2}\bigl\lvert\phi_m\bigr\rangle,
\end{align*}
with
\begin{equation}
    \mathcal{M}_{km} = \sum_{k'=1}^{M'}p_{k'}^\prime\sqrt{p_k p_m}\braket{\phi_k}{\phi_{k'}^\prime}\braket{\phi_{k'}^\prime}{\phi_m},
\end{equation}
where $p_k^{(\prime)}$ and $\phi_k^{(\prime)}$ correspond to the $k$th eigenvalue and eigenvector of $\hat{\rho}^{(\prime)}$. It appears from the above that the total complexity of this evaluation is of order
\begin{equation}
    O(2N^3) + O(2M' N) + O(M^3) + O(MN)\, ,
\end{equation}
where the first term accounts for the eigendecomposition of the two density matrices, the second for the construction of the matrix $\mathcal{M}_{km}$, the third for its diagonalization and the last for the trace. An additional subleading complexity of order $O(n\times N)$ is to be added if the density matrices are obtained via Monte Carlo wave function calculations in order to account for their construction. In contrast, the order $O(2N^3)$ is to be discarded when using the dynamical corner-space method, as the eigendecompositions are known explicitly. Then, for each method, one finally has to leading order and for $M^{(\prime)}\ll N$ the following scaling figures of merit:
\begin{table}[h!]
    \begin{ruledtabular}
    \renewcommand{\arraystretch}{1.2}
    \begin{tabular}{ccc}
         & MCWF & Time-dependent corner space\\
        \hline
       $\mathcal{F}(\hat{\rho},\hat{\rho}')$ & $O(2N^3)$ & $O(\max(M,2M') N)$ \\
       \hline
       $\mathcal{F}(\hat{\rho},\ket{\phi})$ & $O(n N)$ & $O(M N)$ \\
       \hline
       $S(\hat{\rho})$ & $O(N^3)$ & $O(M)$ \\
    \end{tabular}
	\end{ruledtabular}
\end{table}

In practice, the inconvenient scaling of the complexity for the Monte Carlo wave function approach, stemming from the two density-matrix diagonalizations, combined with the necessity of storing dense matrices well beyond the realistically available RAM makes it impossible to compute the fidelity between two mixed states from trajectories for systems larger than $\sim 15$ sites. A similar discussion can be made for the entropy.

\subsection{Scaling of the corner dimension\label{sec:appA.4}}

In our method, the tolerance on the precision of the density matrix is tunable by design and the convergence versus $M$ is ensured for moderately dissipative systems with low entropy. This is contrast to the case of quantum trajectories for which the number of needed trajectories $n$ for a given problem is currently unknown {\it a priori} ~\cite{daley2014traj}. In Fig.~\ref{fig:S1}(b), one can see that for $\gamma T_\mathrm{QFT}=\num{1.42e-2}$, an infidelity below $10^{-3}$ can be reached by choosing $M_{\mathrm{max}} \sim L\ln{L}$. Hence, the corner dimension grows only polynomially with the size $L$, in particular sub-quadratically. Note that this value of $\gamma T_\mathrm{QFT}$ is slightly higher than the state-of-the-art, hence the method is well suited to simulate experimental platforms in the coming years.

\subsection{Integration method and stiffness\label{sec:appA.5}}

\begin{figure*}[t]
    \centering
    \includegraphics[width = 0.825\textwidth]{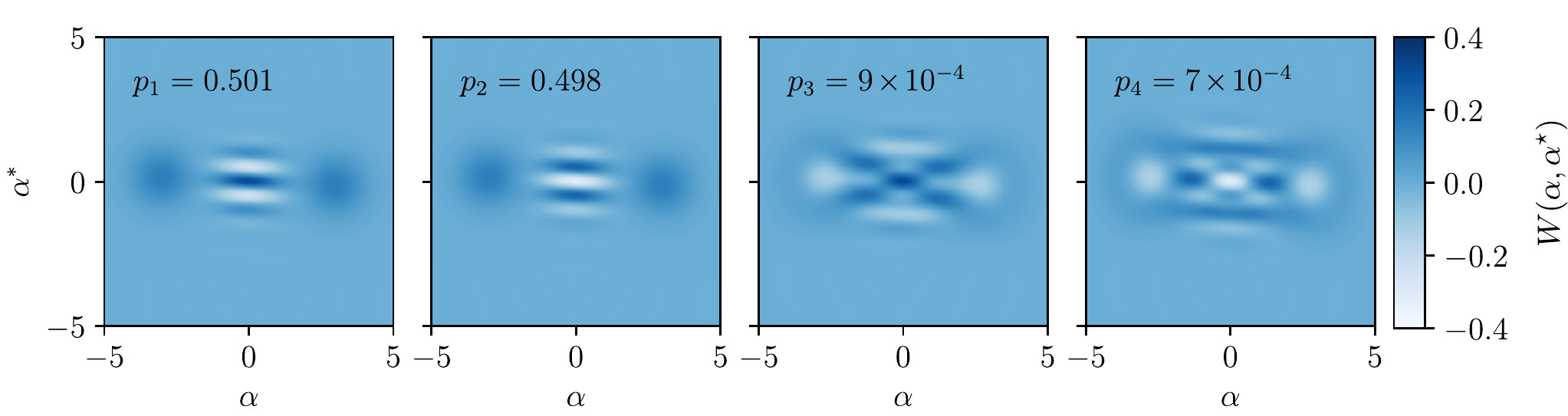}
    \caption{Wigner function of the four leading components of the corner describing a cavity subject to a Lindblad evolution as described in Appendix~\ref{sec:appA.5}, after time $\gamma t = 10$. The first two correspond to states $\ket{C^{\pm}_\alpha}$ with opposite parities. Parameters here are $\kappa/\gamma = 2$, $\omega_c/\gamma = 1$, $K/\gamma = 10$, $G/\gamma = 50$.}
    \label{kerrfig}
\end{figure*}

An issue with the map $\hat{C}(t) \mapsto \hat{C}(t+\delta t)$ that updates the corner basis at each time step is that it involves terms proportional to $\sqrt{\delta t}$ in the Kraus operators. This seems to restrict the method to a first order explicit integration scheme, which could result in poor stability of the method when dealing with stiff dynamics ensued from nonlinear Hamiltonians. However, we circumvent this limitation by separating the pseudo-unitary evolution generated by $\hat{K}_0$ from that of the rest of the Kraus operators. Instead of computing \smash{$T_{im} = \sum_k K_{0,ik}C_{km} = \sum_k(\delta_{ik} - i\delta t \tilde{H}_{ik})C_{km}$}, we perform an exact numerical integration of the differential equation \smash{$\partial_t \hat{C} = -i\tilde{H}\hat{C}$ over the time interval $[t, t+\delta t]$} via an ordinary differential equation (ODE) solver well adapted to the level of stiffness induced by the Hamiltonian. This allows us to treat stiff problems, via adapted implicit methods, and to use adaptive time stepping.

To illustrate this point, let us consider a Kerr-cat qubit~\cite{Grimm_2020} as described by the following Hamiltonian:\begin{equation}
    \hat{H} = K\hat{a}^\dagger\hat{a}^\dagger\hat{a}\hat{a} + \omega_c \hat{a}^\dagger \hat{a} + G(\hat{a}^2+ \hat{a}^{\dagger 2})\, ,
\end{equation} where $\hat{a}$ ($\hat{a}^\dagger$) is a cavity mode annihilation (creation) operator, $\omega_c$ is the frequency of the cavity, $K$ is its Kerr nonlinearity, and $G$ is the two-photon driving frequency. If one considers jump operators $\hat{J}_1 = \sqrt{\gamma} \hat{a}$, $\hat{J}_2 = \sqrt{\kappa} \hat{a}^2$ that describe single- and two-photon losses, respectively, a logical qubit can be conceived by considering logical states $\ket{+} = \ket{C^+_\alpha} = (\ket{\alpha} + \mathrm{i}\ket{- \alpha}) / \sqrt{2}$ and $\ket{-} = \ket{C^-_\alpha} = (\ket{\alpha} - \mathrm{i} \ket{- \alpha}) / \sqrt{2}$ as these are steady states of the system in the $\gamma \rightarrow 0$ limit. Numerically, the simulation of such systems cannot be efficiently treated with explicit first-order methods as the differential equation corresponding to their time evolution is stiff because of the Kerr nonlinearity. Thanks to the numerical integrator used for the coherent part of the evolution of the corner $\hat{C}$, our method is capable of describing such systems, whose entropy is limited when $\kappa \gg \gamma$. In Fig.~\ref{kerrfig}, the Wigner functions of each of the four most populated states of the corner are shown with their associated probabilities $p_i$, after having evolved the system for a time $t = 10/\gamma$ via the corner-space method, setting a photon cutoff $n_\mathrm{ph} = 20$. One sees that the low-dimensional basis found by the corner indeed closely matches that of a qubit with Schrödinger-cat logical states while keeping track of the effects of the single-body loss on the lowly probable states. By tuning the tolerance parameter of the method $\epsilon$, such dissipative effects can be captured to any desired order. Our method therefore enables one to investigate the dynamics of such systems in an efficient way and could be used to understand how the single-photon dissipation impacts quantum operations in multi-Kerr-cat-qubit systems, among other applications involving bosonic qubits.

\section{Continuous-time error model\label{sec:appB}}

In this appendix we discuss the differences between discrete-time and continuous-time error models. A convenient and widely used model for quantum computation is the gate-based model~\cite{nielsen_chuang_2010}. For closed systems, this model is strictly equivalent to the successive application of unitary time-evolution operators of the form \smash{$\hat{U}_{G} = e^{-i\hat{H}_G\tau}$} with \smash{$\hat{H}_G$} as the gate Hamiltonian and $\tau$ as the gate time. In most current classical simulations of noisy quantum processors, errors are accounted for by extending the gate-based model to what has been coined digital error models~\cite{Arute2019}. These consist in applying noise gates after each unitary gate, expressed as Kraus operators, in analogy to error models for classical processors. However, in general, a quantum system is subject to a completely-positive, trace-preserving map acting on the system density matrix $\hat{\rho}$ in continuous time. The generator $\mathcal{L}$ of such a map can always be expressed in Lindblad form as a Liouvillian~\cite{breuer_theory_nodate} whose action on the density matrix takes the form $\mathcal{L}[\hat{\rho}] = \mathcal{U}[\hat{\rho}] + \mathcal{D}[\hat{\rho}]$, with $\mathcal{U}$ and $\mathcal{D}$ as the unitary and dissipative contributions to the time evolution of $\hat{\rho}$. Explicitly, for a quantum gate $G$, one has:
\begin{gather}
    \mathcal{U}[\hat{\rho}] = -\mathrm{i}[\hat{H}_G,\hat{\rho}],\\
    \mathcal{D}[\hat{\rho}] = \sum_{i=1}^D\big(\hat{J}_i\hat{\rho}\hat{J}_i^\dagger - \frac{1}{2}\lbrace\hat{J}_i^\dagger\hat{J}_i,\hat{\rho}\rbrace\big),
\end{gather}
with $\hat{J}_i$ as the jump operators that describe the dissipative channels, which take a simple form when the environment can be treated within the Born-Markov approximation. Given an initial density matrix $\hat{\rho}$, after a time interval $\tau$, the density matrix at the output of the gate $ \hat{\rho}(\tau)$ is given by
\begin{equation}
    \hat{\rho}(\tau) = e^{\tau\mathcal{L}} \hat{\rho}.
\end{equation}
One can always separate the unitary and dissipative parts of $\mathcal{L}$ to obtain
\begin{equation}
    \hat{\rho}(\tau) = \mathcal{E}_G\left[\hat{U}_G(\tau)\hat{\rho}\hat{U}_G^\dagger(\tau)\right] \equiv \mathcal{E}_G(\hat{\rho}_{\hat{U}_G}),
\end{equation}
where we have defined the density matrix after the ideal unitary process \smash{$\hat{\rho}_{\hat{U}_G} = \hat{U}_G(\tau)\hat{\rho}\hat{U}_G^\dagger(\tau)$} with $\hat{U}_G(\tau)$ as the time evolution operator that corresponds to the application of Hamiltonian $\hat{H}_G$ for time $\tau$ and $\mathcal{E}_G$ as a non-unitary map. In cases where $\mathcal{D}$ and $\mathcal{U}$ commute, one explicitly has \smash{$\hat{U}_G(\tau) = e^{-i\hat{H}_G\tau}$}, and \smash{$\mathcal{E}_G = e^{\tau\mathcal{D}}$} for the unitary and the error processes, respectively. However, in general, obtaining $\mathcal{E}_G$ is nontrivial, and experimentally one would need to perform a tomography on each gate to determine its exact error process: this is known as quantum process tomography~\cite{Mohseni_2008}.

Here we numerically simulate the quantum process tomography of a controlled phase gate of duration $\tau = \pi/2\delta$, $\delta$ being the Rabi frequency of the qubits. To do so, we decompose the error process $\mathcal{E}$ in the Pauli basis as
\begin{equation}
    \hat{\rho}(\tau) = \mathcal{E}(\hat{\rho}_{CR_{\pi/2}}) = \sum_{m,n}\chi^\mathrm{err}_{mn}\hat{P}_m \hat{\rho}_{CR_{\pi/2}}\hat{P}_n^{\dagger},
\end{equation}
with $\hat{\rho}_{CR_{\pi/2}}$ as the output density matrix corresponding to an ideal controlled phase gate, $\hat{P}_n \in \{I, X, Y, Z\}^{\otimes 2}$ as the generators of the Pauli group on two qubits, and $\chi^{\mathrm{err}}$ as the error matrix that completely characterizes the error process and that we aim at numerically determining.

\begin{figure*}[t]
    \centering
    \includegraphics[width = \textwidth]{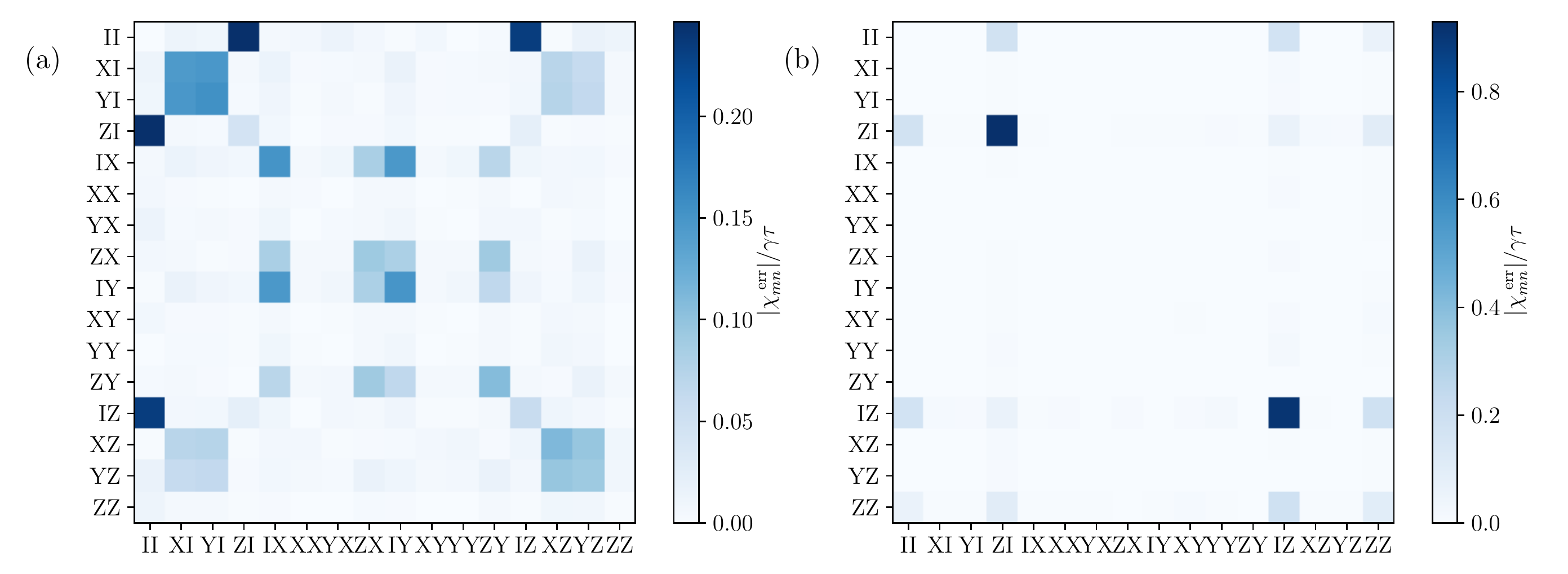}
    \caption{Absolute value of the elements of the error matrices $\chi^\mathrm{err}$ (normalized by the characteristic error magnitude $\gamma\tau$) after the continuous-time evolution described in Appendix~\ref{sec:appB}. In panel (a), the system is subject to dissipation ($\hat{J}_i = \sqrt{\gamma}\sigma^-_i$) and in panel (b) to decoherence channels ($\hat{J}_i = \sqrt{\gamma}\sigma^z_i$). Here $\gamma/\delta = 10^{-3}$. For each error matrix the element $\chi_{11}$ has been set to zero for readability (for such weak values of $\gamma$, $\chi_{11} \sim 1$).}
    \label{fig:S2}
\end{figure*}
In Fig.~\ref{fig:S2}, the error matrix corresponding to a noisy $CR_{\pi/2}$ gate is shown for two different cases. Panel (b) corresponds to a Lindblad evolution with jump operators $\hat{J}_i = \sqrt{\gamma}\hat{\sigma}^-_i$, and panel (a) corresponds to a Lindblad evolution with jump operators $\hat{J}_i = \sqrt{\gamma}\hat{\sigma}^z_i$. By using a Choi decomposition~\cite{obrien_2004}, one can numerically estimate the error matrix that recovers the density matrix $\hat{\rho}(\tau)$ of the realistic Lindblad evolution.
In the first case, $\hat{J}_i = \sqrt{\gamma}\hat{\sigma}^-_i$ was considered, which does not commute with the Hamiltonian. This leads to a complex error process accounting for the spatial propagation of local errors upon (non-local) Hamiltonian time evolution. In this case, this manifests through the presence of single- and two-qubit error events with different magnitudes in the components of the error matrix. Similar error matrices have also been found experimentally~\cite{weinstein_quantum_2004}. Each non-negligible element corresponds to a Kraus operator to be applied after the ideal gate in the digital-error approach. Although two local jump operators suffice to describe the qubit-environment interaction, up to $16$ Kraus operators could be necessary in such an approach. Note that the noise process acting on two qubits at the same time generally has a smaller magnitude, which can justify that one can neglect them for low-depth circuits (the quantum Fourier transform has $L(L+1)/2$ gates, meaning these errors can accumulate in the long run). By contrast, in panel (b) pure-dephasing jump operators were considered, \smash{$\hat{J}_i = \sqrt{\gamma}\hat{\sigma}^z_i$}, which commute with the controlled phase Hamiltonian. Hence, the error process only involves $Z$ errors in this case, although not strictly local (note the $ZZ$ component).

As appears from the above discussion, modeling errors in continuous time is of fundamental and applied importance for the following reasons:
\begin{enumerate}[(1)]
    \item The errors induced by the presence of local dissipative events as captured by local jump operators cannot be accounted for via local noise gates, and are affected by the applied Hamiltonian, an effect that is generally not described by digital error models. For $d$-qubit gates, the number of noise gates to be applied goes up to $4^d$ in a digital-error approach.
    \item For circuits, such as the quantum Fourier transform, there are $L$ controlled rotation gates that are applied for different times, hence one would need to perform tomography on each of them to recover their error matrices (as well as for the Hadamard gate). For multi-qubit gates such as the Toffoli gate, tomography becomes even more expensive.
    \item Characterizing the jump operators of an experimental platform is much easier than performing tomography to obtain the error process for each gate. For example, in the case of superconducting circuits,  measuring the $T_1$ and $T_2$ relaxation times is enough.
    \item Our method is also able to treat collective dissipative processes described by jump operators, such as $\hat{J}_z = (1/\sqrt{L}) \sum_i \hat{\sigma}^z_i$~\cite{nissen2013,Shammah_2018}. This corresponds to a single Kraus operator in our method and is, hence, inexpensive. With a digital error model, describing collective effects would require to perform tomography on a $L$-qubit system, which is intractable since it requires $O(4^L)$ measurements.
\end{enumerate}

\bibliography{bib}

\end{document}